# Electrically controlled heat transport in multilayer graphene

Pietro Steiner[1,2], Saqeeb Adnan[3], Muhammed Said Ergoktas[1,2], Julien Barrier[2,4], Xiaoxiao Yu[1,2], Vicente Orts[1,2], Gokhan Bakan[1,2], Jonathan Aze[1,2], Yury Malevich[1,2], Kaiyuan Wang[1,2], Pietro Cataldi[1,2,5], Mark Bisset[2], Sinan Balci[6], Sefik Suzer[7], Marat Khafizov[3], Coskun Kocabas[1,2,8]

1. Department of Materials, University of Manchester, Manchester, UK
2. National Graphene Institute, University of Manchester, Manchester, UK
3. Department of Mechanical and Aerospace Engineering, The Ohio State University, Columbus, USA
4. Department of Physics and Astronomy, University of Manchester, Manchester, UK
5. Center for Nano Science and Technology @PoliMi, Istituto Italiano di Tecnologia, Milano, Italy
6. Department of Photonics, Izmir Institute of Technology, Izmir, Turkey
7. Department of Chemistry, Bilkent University, Ankara Turkey
8. Henry Royce Institute for Advanced Materials, University of Manchester, Manchester, UK

Corresponding author: coskun.kocabas@manchester.ac.uk

**The ability to control heat transport with electrical signals has been an outstanding challenge due to the lack of efficient electrothermal materials. Previous attempts have mainly concentrated on phase-change and layered materials and encountered various problems such as low thermal conductivities and modest on/off ratios. Here, we demonstrate a graphene-based electrothermal switch enabling electrically tuneable heat flow. The device uses reversible electro-intercalation of ions to modulate the in-plane thermal conductivity of graphene film by over thirteen-fold *via* electrically tuneable phonon scattering. We anticipate that our results could provide a realistic pathway for adaptive thermal transport enabling a new class of electrically driven thermal devices which would find a broad spectrum of applications in aerospace and microelectronics.**

Electrical control of the thermal conductivity of materials could lead to various innovative applications[1–3], such as reconfigurable thermal management for space applications[4], smart heat-dissipating materials enabling steering of the heat flow in the desired directions [5,6] and electrically driven thermal circuits[3]. These applications, however, require an active device i.e. a thermal switch, which can be reversibly changed between an off-state (high thermal conduction) to an on-state (low thermal conduction) on-demand[1], mimicking the functionality of a transistor in electronics. This thermal action can be achieved by altering the lattice ($k_p$) or electronic ($k_e$) contribution to the overall thermal conduction ($k_T$)[7]. Literature includes various approaches to control these contributions. The electrostatic gating of 2D-materials is used to control the electronic contribution[8]. Electrical switching of domains in ferroelectric materials and doping induced phase-transitions in perovskites[2] and layered materials[9] have been investigated to control the lattice contribution. As an alternative, mechanical means have been used to alter the form factor to achieve tuneable thermal conduction[10,11] (see the benchmarking in Table S1[1,12]). Although the previously demonstrated devices provide some variation of the out of plane thermal conductivity in the active layer, the thermal conductance of the devices stays constant due to the small contribution of the thin active layer to the overall thermal conduction of the device[13–15]. The remaining challenge is to achieve a large modulation of in-plane thermal conductance in a scalable device layout. This paper introduces a practical realisation of electrothermal switches using graphene-based devices.

To overcome this challenge, we have designed a device utilising the very high in-plane thermal conductivity (>1000 W m$^{-1}$ K$^{-1}$) of multilayer graphene film (~100 nm) on ultra-thin (~5 µm) low thermal conductivity polymer membranes (~1.1 W m$^{-1}$ K$^{-1}$), soaked with an ionic liquid electrolyte (DEME-TFSI). The fabrication scheme exploits lamination of the thin polymer membrane on CVD grown graphene film and deposition of platinum electrodes (~50 nm) on the back of the membrane (Fig. 1a). The device utilises reversible intercalation of ions to modulate the in-plane thermal conductivity of the multilayer graphene layer. The thickness of the graphene layer (100nm ~ 300 layers) is simultaneously optimised to achieve efficient intercalation[16,17] with large thermal conductance. Figure 1b shows the thermal circuit diagram of the device, including the two parallel thermal resistors associated with the active graphene layer and the passive substrate. Using an ultrathin membrane reduces the thermal conductance of the substrate ($\sigma_{sub} = \kappa_{sub}\, t_{sub} = \frac{1}{R_{sub}}$) which must be smaller than the thermal conductance of the graphene layer ($\sigma_{MLG} = \kappa_{MLG}\, t_{MLG} = \frac{1}{R_{MLG}}$), such that the modulation yields an observable change in the effective conductance of the whole device ($\sigma_T = \frac{1}{R_{MLG}} + \frac{1}{R_{sub}}$). The thermal conductivity of the graphene film is controlled by the reversable intercalation of ions that can be characterised by the intercalation stage, n indicating number of graphene layers between two intercalated layers (Fig. 1c). The effective distance between periodic repeat of intercalant layer can be written as $d = (n-1)d_0 + d_i$ where $d_0$ and $d_i$ represents Van der Waals distance of graphene and intercalated layers. This electrically tuneable structure enable reversable phonon scattering mainly through four mechanism, (1) scattering from point defects due to intercalant ions, (2) scattering from boundaries between intercalated and unintercalated layers and (3) Umklapp scattering leading to intrinsic reduction of conductivity. These mechanisms reduce the phonon contributions, however, (4) the electronic contribution increases with the doping and becomes the dominant heat conduction mechanism at the intercalated state [18]. We first present the thermal and spectroscopic characterisation of the thermal switches, then compare these results with an analytical thermal model. We have also analysed the contribution of the out-of-plane thermal conduction on our measurements.

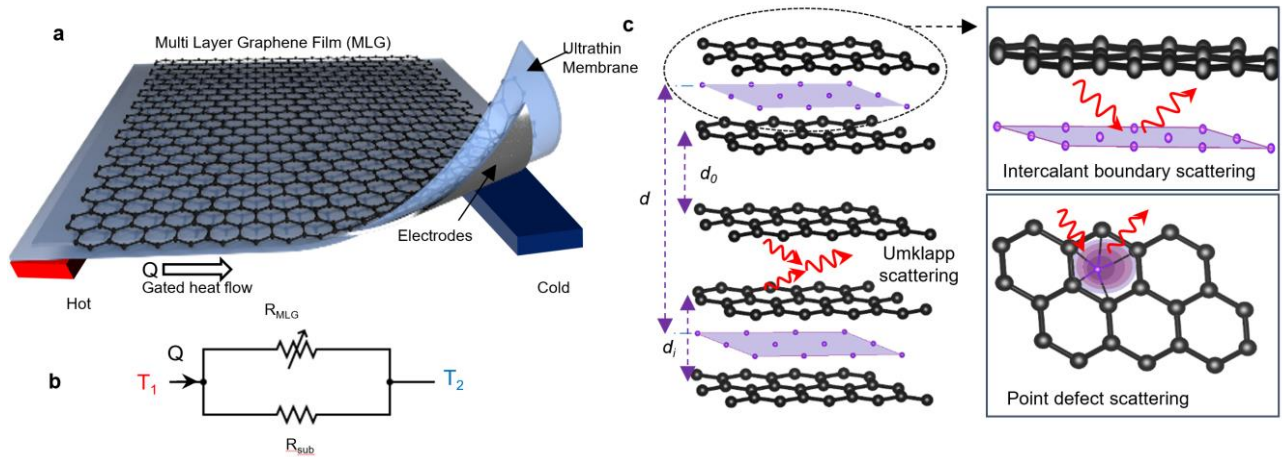

**Figure 1. The structure and operation principle of the device**: **a,** Schematic of the thermal switch consisting of multilayer graphene film on a 5-µm thick polyethylene membrane soaked with ionic liquid electrolyte and platinum counter electrode at the back. **b,** Thermal circuit model of the device. The thermal conduction through the graphene layer and the substrate is represented with a variable and a fixed thermal resistor. **c,** Representation of possible phonon scattering mechanism including Umklapp scattering, intercalation induced point defects, and scattering from two types of boundaries.

Figure 2a shows the fabricated device consisting of two separate back electrodes, enabling controlled intercalation on the selected area (Fig.2b). The process can be visualised with an infrared camera due to the suppressed IR emissivity of the intercalated graphene film[15] (Fig. 2c). To measure the thermal conductivity of the overall device and the active layer, we implemented infrared thermography[19] and modulated thermoreflectance methods (Fig. 2d). These techniques measure the phase delay of thermal waves in millimetres (Fig. 2e) and micrometres (Fig. 2g) length scales, providing complementary information for the device and active graphene layer. Using the continuum-based thermal heat diffusion model[20–22], we were able to extract the extrinsic and intrinsic thermal conductivity from the measured thermal wave profile in the device. This model solves the anisotropic heat diffusion equation across a series of layers in a stacked multilayer system[20]. We compared this solution to the experimental thermal wave profiles with a least square minimisation routine while using in-plane and out of plane thermal conductivity as a fitting parameter. Additional details are reported in the supplementary information section 2.

In Figure 2f, we show the variation of actual thermal conductivity of the device with the applied voltage. Although the electronic contribution of the thermal conductivity increases by 30 fold, we observed a sudden decrease in the thermal conductivity after -2.5V indicating an enhanced phonon scattering. The reduction of in-plane thermal conductivity was further confirmed with the thermoreflectance measurements (at a modulation frequency of 1 kHz), showing that the thermal conductivity modulation of graphene film varies from ~1100 down to 80 W $m^{-1}$ $K^{-1}$, providing a tuneable thermal conductivity of the overall device between ~35 down to 10 W $m^{-1}$ $K^{-1}$. After the first cycle, the thermal conductivity was recovered to over 95% of its initial value when the voltage was brought to +2V to fully deintercalate the film (Fig. S7). The enhanced phonon scattering in the intercalated graphene film can be attributed to a boundary scattering from intercalant - graphene layer interfaces and defects resulting from the graphene's carbon bonds ionically interacting with the dangling bonds of the intercalant ions. These ionic interactions are much stronger than the weak van der Waals interaction between the graphene layers resulting in large scattering rates due to charged impurities.

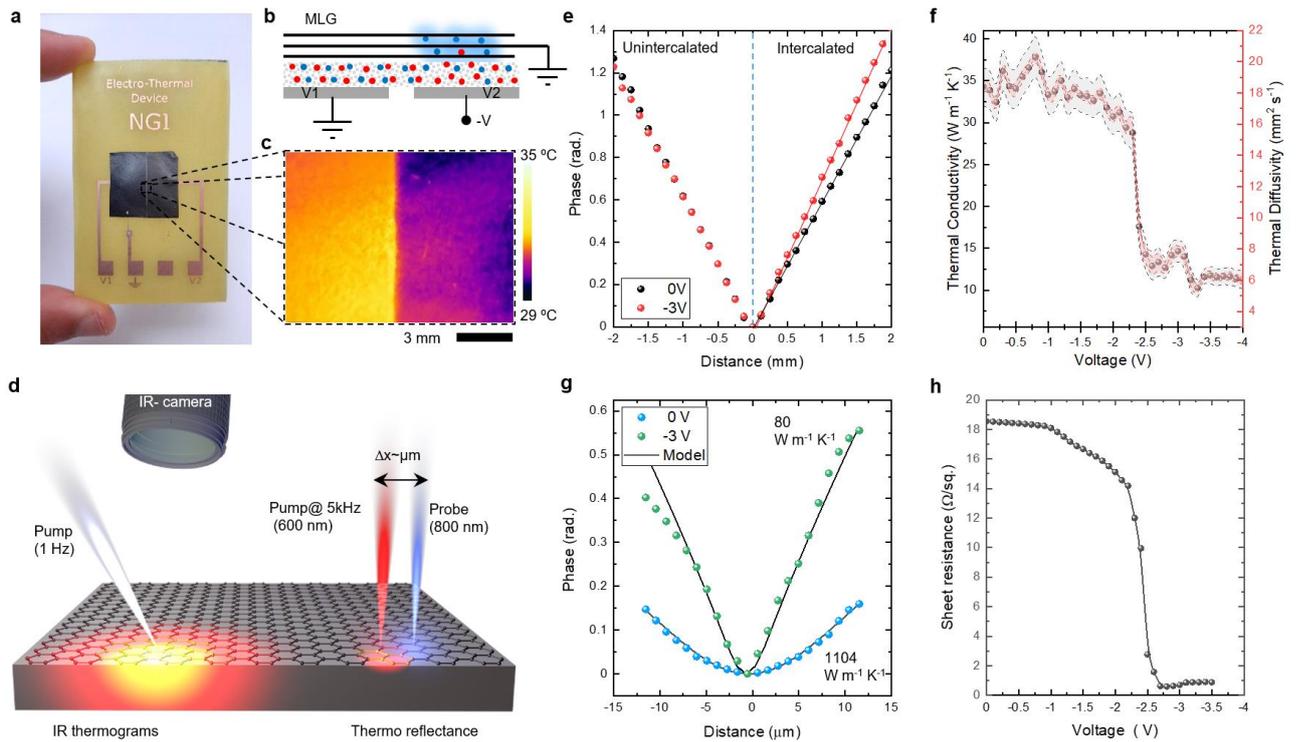

**Figure 2. Characterisation of the thermal switch**: **a,** Photograph of the fabricated device wired on a PCB board with 12 mm hole at the centre to suspend the device. The two back electrodes enabled area-selective intercalation. **b,** Cross-section and **c,** the thermal image of the device showing that only half of the device is intercalated for $V_1=0$ V and $V_2=-3$V. The unintercalated area was used as a reference for the measurements. **d,** Schematic showing the thermo-reflectance and IR thermograms for thermal characterisation in micrometre and millimetre length scales. For thermo-reflectance measurements, the 600nm pump laser is modulated at 5kHz, and the thermal wave profile is measured by moving the 800nm probe laser. IR thermography uses a camera to record the propagation of heat waves generated by a supercontinuum laser modulated at 1 Hz. **e,** Measured phase delay of the heat waves as a function of distance from the laser spot. The change in the slope of phase for the intercalated area indicates the modulation of the thermal diffusivity D, which is calculated as $D=\omega r^2/2\phi^2$ where $\omega$ is the modulation frequency, r is the distance from the source, and $\phi$ is the phase delay. The thermal diffusivity was subsequently converted into conductivity via the independent measurement of the density and the specific heat capacity. **f,** Variation of the thermal conductivity (black) and diffusivity (red) as a function of the device voltage displaying a three-fold decrease. **g,** Measured phase delay recorded by the thermoreflectance measurements and the fitting results of the thermoreflectance model at 0 and -3V. **h,** Voltage dependence of the sheet resistance of graphene layer measured using 4-point probe method.

To elucidate the intercalation process further, we performed *in-situ* spectroscopic characterisations. In Figure 3a, we show the evolution of the Raman spectra of the multilayer graphene film under the bias voltage. We observed a blue shift in the frequency and splitting of the G-band at higher voltages. The splitting of the G-band indeed indicates partial intercalation of the multilayer film. The $G_2$ band is associated with the interfacial mode between intercalated and unintercalated layers. We also observed suppression of the 2D band due to Pauli blocking, indicating a large Fermi energy shift of > 0.8 eV obtained from the condition of blocking interband transitions as $2E_F > E_{ex} - 2\hbar\Omega_D$ where $E_F$ is the Fermi energy, $E_{ex}$ (1.96 eV) is the energy of the excitation laser and $\hbar\Omega_D$ (~0.17 eV) is the energy of D-band phonons[23] (Fig. S8). It should be emphasised here that we did not observe a significant defect band (1350 cm$^{-1}$), indicating that the intercalation process is not detrimental to the quality of the graphene film. To measure the voltage dependent intercalation stage, we performed in situ X-ray diffraction (XRD) on the device (Fig. 3b). When we apply voltage >2V, the (002) peak of graphite gradually disappears and higher order diffraction peaks due to the periodic repeat of intercalant layer appear. We estimated the intercalation stage from the ratio of the peak positions of $00(n+2)$ and $00(n+1)$[24,25]. From these results, we were able to obtain voltage dependent intercalation stages (Fig. 3c, Fig. S9). We have supported these results with *in-situ* X-ray photoelectron spectroscopy providing chemically specific information for the intercalation process(Fig. 10). The ionic liquid contains two nitrogen atoms, one with a positive charge (quaternized nitrogen, N$^+$) and the other with a negative charge (imide nitrogen, N$^-$), which yield two well resolved N1s peaks. At 0V we observed symmetric N1s peaks indicating no net charge on the graphene layer, however at -3.0 V, there is a significant charge imbalance ~20% (the ratio of to N$^-$ to N$^+$) which is responsible for electrostatic doping on the graphene layers. XPS also provide a direct measurement of the Fermi energy. Since the graphene layer is grounded, the shift in the binding energy is due to the Fermi energy change. The binding energy of N$^+$ decreases by 0.7 eV at -3V which is consistent with p-type doping and the value obtained from Raman measurements.

On the basis of these observations, we present an analytical model to get some insight on the mechanism impacting thermal conductivity during the intercalation process. We employ an empirical Klemens-Callaway model for thermal conductivity previously parametrised for graphite[22,26,27]. The model is based on the kinetic theory of phonon gas expression obtained from the relaxation time approximation of Boltzmann transport $\kappa = \sum_q C|\vec{v}\cdot\hat{n}|^2\tau$ [21]. The summation is over all the phonon modes within the Brillouin zone, *C* is specific heat, $\vec{v}$ is phonon velocity, $\hat{n}$ is direction along which thermal conductivity is calculated, and $\tau$ is phonon lifetime which considers Umklapp scattering leading to intrinsic reduction of conductivity, intercalation induced defects acting as point defects, and scattering from two types of boundaries. In-plane mean free path of phonons is limited by the grain size and cross-plane boundary determined by the intercalation stage obtained from XRD results, that is the number of graphene layers between successive intercalant layers. Umklapp and point defect scattering take their usual form applicable to 2D materials[26]. The boundary scattering assuming diffuse limit is given by $\tau^{-1} = 2\vec{v} \cdot \hat{n}_{\parallel}/D_g + 2\vec{v} \cdot \hat{n}_{\perp}/d$, where $D_g$ is grain size in-plane direction and $d$ is cross-plane thickness determined by the intercalation stage, and are unit vectors for in-plane $\hat{n}_{\parallel}$ and cross-plane directions $\hat{n}_{\perp}$, respectively. For simplicity, all the phonons are described using Debye expression $\omega = vq$, where the velocities are obtained from the solution of Christoffel equation[21].

To model conductivity as a function of applied voltage, we take into account the microstructural changes taking place under the external stressor[28]. The cross-plane thickness was determined based on intercalation stage obtained from XRD data. While we didn't determine the absolute concentration of point defects, we assumed that it is proportional to in-plane electrical conductivity obtained from sheet resistance measurement, assuming that contribution of charged impurity defects to electrical conductivity and reduction of thermal conductivity are related. When considered independently scattering by point defect and intercalation delamination provide qualitatively similar trends in thermal conductivity modulation obtained from thermal reflectance and IR thermogram experiments (Fig. 3d). Sensitivity of the laterally resolved thermal wave imaging approaches to basal plane conductivity of graphene offers a significant advantage over previous reports. Considering the range of cross-plane conductivities predicted by the anisotropic thermal conductivity model we estimate the uncertainty in measured basal plane conductivity. We obtain values of 1105 ± 50 W m$^{-1}$ K$^{-1}$ for in-plane assuming 6.8 ± 3.5 W m$^{-1}$ K$^{-1}$ for cross-plane[7] for pristine. Similarly, for intercalated sample we obtain 80 ± 5 W m$^{-1}$ K$^{-1}$ for in-plane assuming 1.5 ± 1 W m$^{-1}$ K$^{-1}$ for cross-plane conductivity. For detailed explanation of the model see the supplementary information section 6.

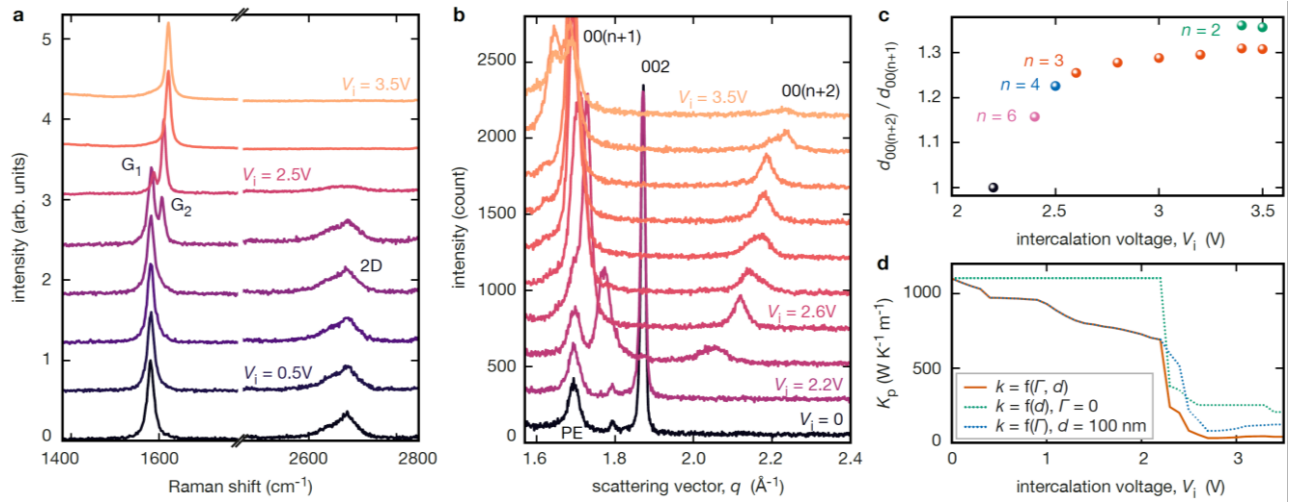

**Figure 3. Spectroscopic characterisation of the thermal switches**. **a,** *In situ* Raman spectrum of the graphene layer recorded at different bias voltages, showing the variation of the frequency and intensity of G and 2D bands. **b,** Voltage dependent *in situ* XRD showing the characteristic graphite peaks at 0V and appearance of higher order diffraction peaks 00(n+1) and 00(n+2) due to the periodic repeat of intercalant layer appears. **c,** Ratio of peak positions 00(n+2) / 00(n+1) as a function of the voltage, highlighted in different colours the dominant graphite intercalation stages, n. Two dominant peaks are coexisting when a voltage of 3.4V and 3.5V is applied (Fig.S9). **d,** Predicted thermal conductivity profile of the MLG sheet as a function of bias voltage using the anisotropic thermal conductivity model. The brown line represents the model accounting for both the defect scattering strength ($\Gamma$) and the boundary scattering distance (d). The dotted lines displayed the cases where only one of those two effects are considered. The model is fitted for the experimentally measured (MTR) thermal conductivity values at 0 and 2.5V.

Scanning thermal microscopy (SThM) can add additional insight to the detailed features of these devices at the nanoscopic scale[10,29]. We used an atomic force microscope (AFM) tip with a resistive element near the apex to heat the sample via Joule heating and simultaneously monitored its temperature. In this experiments we used highly orientated pyrolytic graphite flake (HOPG) exfoliated onto a porous polyethylene membrane. A change in the thermal conductivity of the sample or their temperature produces a variation in the tip resistance, providing qualitatively thermal information of the specimen (Fig. 4a). In Figure 4b and 4d we show the correlation between the AFM height image and thermal image of the boundary between the graphene film and the membrane. Due to the low thermal conductivity of the membrane, the thermal image shows higher voltages indicating low thermal dissipation at nanometre length scale. We performed in-situ SThM to monitor the spatial variation of the intercalation process and the suppression of the local thermal conductivity (Fig. 4d). It is clear that high-resolution thermal contrast can be obtained in the SThM images for the intercalated regions, which gradually diffuse with the applied voltage. The area with thicker graphene layers shows slower intercalation process thus delayed thermal switching. The SThM maps display an inhomogeneity of the intercalation process, highlighting the grain boundaries and defects that initiates the process. Figure 4e and 4f show the distribution of the SThM signal and its variation with the applied voltage. These SThM images show that the intercalation process is initiated from the defects at the grain boundaries on the graphene film then diffuses in the plane (Fig. S14).

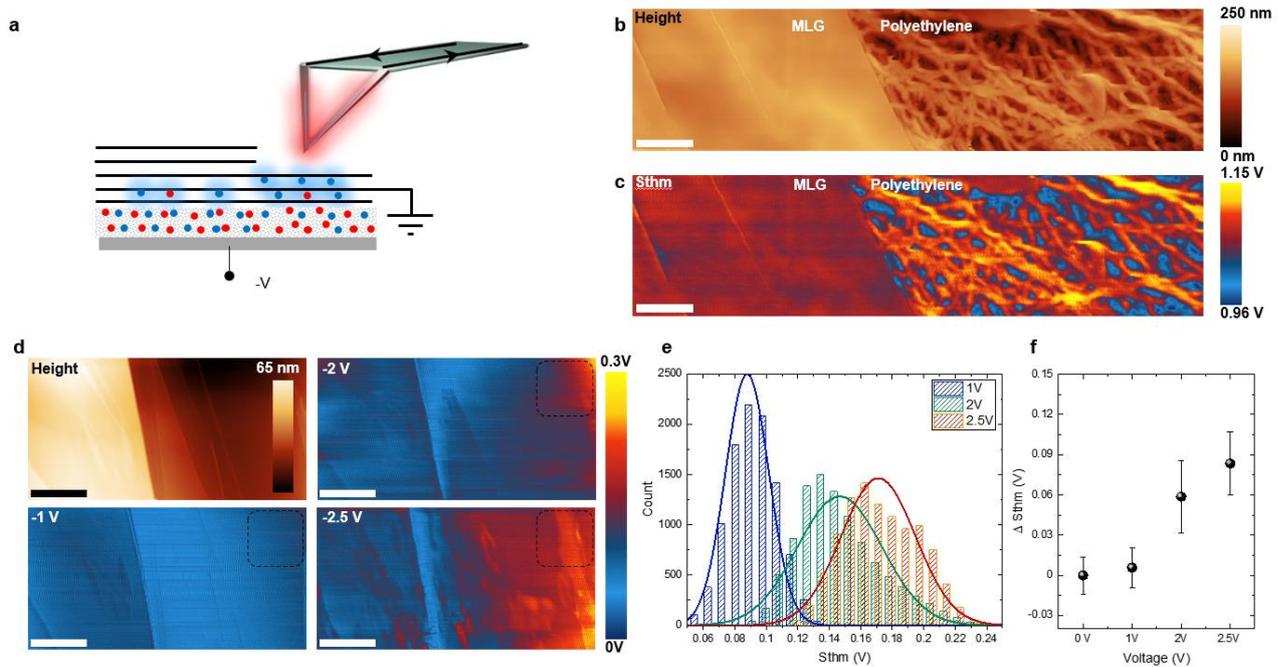

**Figure 4. Nanoscale visualisation of thermal switching: a,** Schematic showing the working principle of the scanning thermal microscopy. The joule heated AFM tip produced a local increase of temperature beneath the tip apex. Measuring the heat dissipation at different locations enabled us to estimate the thermal conductivity nature of the specimen qualitatively. **b,c,** Topography and SThM maps of highly orientated pyrolytic graphite flake (HOPG) onto porous polyethene, respectively (scale bar 1μm). **d,** Height and SThM signal at different device voltages. The height map displayed a sharp step of approximately 25 nm. The intercalation process started from the right corner at 2V and subsequently propagated throughout the all thinner side of the samples at 2.5V. **e,** Histogram extracted from the dotted square (1μm$^2$) in the right side of the SThM maps showing the gradual expansion of the intercalated area. **f,** Graph shows the shift of mean thermal voltage against the bias voltage. The measurement error was calculated by calculating the histogram's standard deviation.

To showcase the promises of our approach, we demonstrate the steering of heat waves by programmable in-plane thermal diffusivity. Figure 5a shows the fabricated device and its layout consisting of a continuous graphene film with 10 independent back electrodes (Fig. 5b) shaped into equal slices. The area selective intercalation of graphene film enables reconfigurable in-plane anisotropy, as shown in Figure 5c for three different device configurations. To generate the heat waves, we focused the excitation laser on the centre point and measured the dissipation of heat while applying different voltages on the back electrodes (Video S1). Such configuration generates anisotropic thermal diffusivity guiding more heat flow in high diffusivity regions (unintercalated area). We observed an increase in the temperature on the high diffusivity side shown in Figure 5e. The temperature change can be enhanced by designing the geometry of the back electrodes (Fig. S17). This observation suggests that the programmable in-plane thermal diffusivity can guide the in-plane heat diffusion defined by the voltages configuration on the electrodes.

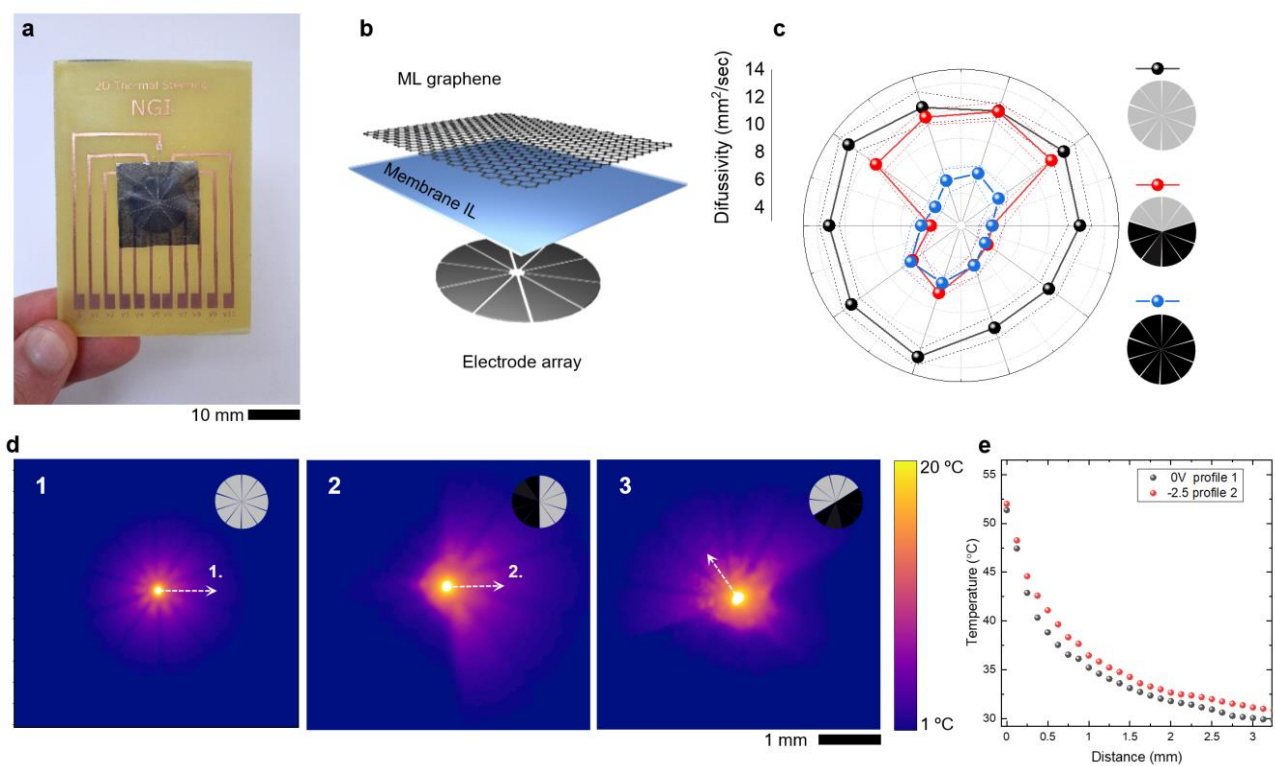

**Figure 5. Programmable 2D steering of heat**. **a,** Optical image of the fabricated 2D thermal steering device, mounted onto a custom-build PCB. **b,** The device structure of 2D thermal steering device: at the top a multilayer graphene electrode, a porous PE substrate soaked with ionic liquid, and ten independent back electrodes. **c,** In-plane radial thermal diffusivity modulation for three specific configurations: When the device is not intercalated (black dots), when over half of the device is intercalated (red dots) and when the device is fully intercalated (blue dots). **d,** Three long-range IR thermograms are reported for different intercalation configurations, highlighting the device's steering capability. **e,** The modulation of thermal conduction as a function of the direction joined with the reversible intercalation process enables the preferential heat dissipation in a specific direction, producing an increase in temperature along the unintercalated region.

In summary, our results demonstrate that the electrical controllability of thermal conductivity of graphene films enables a practical realisation of electrothermal switches, outperforming the existing literature (benchmarking studies, SI section 1). We anticipate that these proof-of-concept devices may pave a realistic pathway for a new class of active thermal devices which can be utilised for energy harvesting, thermal management or thermal circuits. Additional to the tunable thermal conductivity, these devices yield infrared emissivity modulation[10–14]. This dual functionality would enable more flexibility for designing adaptive thermal management systems for space applications.

**Acknowledgements:** This research is supported by Airbus, EPSRC funding Centre for Doctoral Training (CDT) Graphene-NOWNANO and European Research Council through ERC-Consolidator Grant (grant no 682723, SmartGraphene). We also acknowledge Henry Royce Institute for Advanced Materials for the NanoIR characterisation facility. S.A. and M.K. acknowledge support from the Center for Thermal Energy Transport under Irradiation, an Energy Frontier Research Center funded by the US Department of Energy, Office of Science, Office of Basic Energy Science.

**Author contributions:** P.S. and C.K. conceived the idea and designed the experiments. P.S. fabricated the devices and performed IR thermograms thermal characterisation experiments, SThM measurements. S.A. and M.K. performed thermo reflectance experiments and developed the thermal model. S.B. synthesised the graphene samples. S.S. performed the XPS characterisation. P.S. performed the electrical characterisation measurements. V.H performed DSC measurements. M.E performed Raman measurements. P.S. and X.Y. performed XRD measurements, J.B. performed the XRD data analysis. P.S. and J.A. performed UV-Vis measurements. All authors discussed the results and contributed to the scientific interpretation as well as to the writing of the manuscript.